\documentclass[10pt, aps, prb, twocolumn, superscriptaddress]{revtex4-2}

\usepackage{graphicx}
\usepackage{amsmath, amsfonts, amssymb}
\usepackage{xcolor}
\usepackage{comment}
\usepackage[normalem]{ulem}
\usepackage[hidelinks]{hyperref}

\begin{document}

\title{Non-Bloch band theory of boundary-controlled magnon edge modes in an antiferromagnetic chain}
\author{Suman Debnath}
\affiliation{Department of Physics, Indian Institute of Technology Kanpur, Kalyanpur, Uttar Pradesh 208016, India}

\author{Sonu Verma}
\affiliation{ Institute for High Pressure, Department of Physics, Hanyang University, Seoul 04763, Republic of Korea}

\author{Rohit Mukherjee}
\affiliation{Institute for Theoretical Physics, University of Cologne, Zülpicher Straße 77, 50937 Cologne, Germany}

\author{Arijit Kundu}
\affiliation{Department of Physics, Indian Institute of Technology Kanpur, Kalyanpur, Uttar Pradesh 208016, India}

\begin{abstract}
We define a winding number within the Non-Bloch band theory framework that captures the emergence of magnon edge modes in a one-dimensional antiferromagnetic spin chain, even when the conventional Bloch winding number is trivial. Within linear spin-wave theory, magnon excitations are governed by a non-Hermitian dynamic matrix, despite the underlying Hamiltonian being Hermitian. The symmetry classification of this matrix yields a trivial bulk invariant, however, finite systems exhibit boundary-localized modes, signaling a breakdown of the conventional bulk-boundary correspondence. We further show that these edge modes can be controlled via boundary perturbations. By tuning the boundary potential, the modes can be driven into or out of the bulk spectrum. To resolve the bulk-boundary mismatch, we develop a non-Bloch framework based on a generalized Brillouin zone and a winding number that correctly predicts the presence of edge states. Our results establish boundary-controlled topological transitions that are experimentally accessible through local Zeeman fields or modified edge anisotropy in antiferromagnetic van der Waals nanostructures.
\end{abstract}

\maketitle

\section{Introduction}
Long-range magnetic order in atomically thin two-dimensional systems has long remained a central question in condensed matter physics~\cite{Mermin1966, Hohenberg1967, Park2026, Kumar_2026}. The recent discovery of intrinsic magnetism in van der Waals (vdW) materials has provided a platform to explore this question~\cite{Gong2017, Huang2017, Wang2022, coraux2025}. Early experiments show that monolayer CrI$_3$ exhibits Ising-type ferromagnetic order with a Curie temperature of approximately 45 K, demonstrating that magnetic order can persist down to the two-dimensional limit~\cite{Huang2017}. Subsequent experiments on related vdW magnets, including CrCl$_3$~\cite{McGuire2017}, CrBr$_3$~\cite{Yang2023CrBr3}, CrSBr~\cite{LopezPaz2022}, and transition-metal phosphorus trichalcogenides MPS$_3$ (M = Fe, Ni)~\cite{Lee2016,Kim2019,Burch2018}, have revealed a rich variety of layer-dependent magnetic phases and tunable exchange interactions. In these systems, inequivalent exchange paths arising from structural distortions, strain, or finite-size effects can naturally lead to effective low-dimensional descriptions, including one-dimensional spin models with modulated exchange couplings. An effective quasi-one-dimensional structure can also emerge along the layer stacking direction. Layered magnets such as CrSBr, and certain few-layer CrI$_3$ devices, combine strong intralayer ferromagnetic exchange with weaker interlayer coupling. In regimes where each layer behaves approximately as a macrospin, the stacking direction can be modeled as an effective one-dimensional spin chain~\cite{stetzuhn2025}. Due to structural complexities, various magnetic interactions such as Dzyaloshinskii--Moriya interactions (DMI) and easy-axis anisotropy can be influenced by external perturbations such as strain~\cite{Pizzochero2020}, twist~\cite{Xu2021}, and electric and magnetic fields~\cite{Jiang2018, Badola2026, stetzuhn2025}. These considerations motivate the study of minimal spin models with various interactions that describe low-dimensional magnetic excitations.

A distinctive feature of antiferromagnetic spin waves is that their quadratic Hamiltonian generally takes a bosonic Bogoliubov--de~Gennes (BdG) form~\cite{Colpa1978, Shindou2013a, Peano2018}. The corresponding magnon eigenvalue problem is governed not by the Hermitian Hamiltonian itself, but by the bosonic dynamic matrix. This matrix is generically non-Hermitian, despite the underlying spin Hamiltonian being Hermitian, and is constrained by pseudo-Hermiticity. A real positive-frequency magnon spectrum is tied to the dynamical stability of the assumed magnetic ground state.~\cite{Shindou2013a,McDonald2018,Kondo2020,Kawabata2019}. Thus, magnon BdG systems offer an intrinsic route to non-Hermitian band structures without coupling to reservoirs, parametric driving, or engineered gain and loss.

This observation connects antiferromagnetic magnons to the broader framework of non-Hermitian lattice physics. In non-Hermitian systems, topological invariants evaluated on the ordinary Brillouin zone can fail to predict the boundary spectrum of a finite system, leading to a breakdown of the conventional bulk-boundary correspondence~\cite{Lee2016a, Yao2018,Kunst2018, Okuma2020}. Non-Bloch band theory resolves this problem by continuing the crystal momentum into the complex plane and defining topological invariants on a generalized Brillouin zone (GBZ)~\cite{Yokomizo2019, Yang2020, Yokomizo2021, Bergholtz2021, Ashida_2020}. While pseudo-Hermitian bosonic BdG systems have been classified and explored in several contexts~\cite{McDonald2018,Kondo2020,Slim2024,Busnaina2024}, microscopic magnonic realizations of non-Bloch topology remain comparatively unexplored. In particular, it is important to determine whether the pseudo-Hermiticity inherent to antiferromagnetic spin-wave theory can produce boundary spectra that are invisible to conventional Bloch invariants.

In this work, we address this question using a one-dimensional antiferromagnetic chain with dimerized exchange, next-nearest-neighbor DMI, and easy-axis anisotropy. Within linear spin-wave theory, we show that the conventional Bloch winding number vanishes, while finite chains under open boundary conditions support magnon edge modes. This establishes a concrete mismatch between the Bloch topological characterization and the boundary spectrum of the magnon BdG dynamic matrix. We resolve this mismatch by introducing generalized boundary conditions through tunable onsite potentials at the two ends of the chain. This is motivated by the fact that the abrupt termination of the lattice at open boundaries breaks the exchange bonds at the edge sites, naturally modifying their local onsite energies relative to the bulk. In the analytically tractable limit of zero DMI, we construct the GBZ explicitly and define a non-Bloch winding number on this contour. The invariant takes a quantized value when the edge modes are present and drops to zero when they merge into the bulk continuum. We further show that moderate DMI shifts the magnon spectrum without destroying the boundary modes, up to the instability threshold beyond which the collinear N\'eel state is no longer stable. Our results provide a microscopic magnonic realization of non-Bloch topology in a conservative bosonic system. The generalized boundary potential may be implemented through local Zeeman fields or modified edge anisotropy, allowing the edge modes to be driven into and out of the bulk spectrum. The resulting boundary-controlled topological transitions offer experimentally testable signatures of pseudo-Hermitian non-Bloch physics in antiferromagnetic vdW nanostructures.

The rest of the manuscript is organized as follows. In Sec.~\ref{sec: Model Hamiltonian}, we introduce the microscopic model for the one-dimensional antiferromagnetic chain. Here, we develop the linear spin wave description and discuss the stability of the N\'eel ground state. Sec.~\ref{sec: Finite System with Open Boundary} analyzes the finite system under open boundary conditions (OBC), where we demonstrate the emergence of sublattice-polarized edge modes. In Sec.~\ref{sec: Bulk Winding Number}, we highlight the failure of the conventional bulk winding number to capture these states. To understand this discrepancy, Sec.~\ref{sec: Boundary Perturbation and Tunable Edge Modes} explores boundary perturbations and the resulting tunable edge modes. In Sec.~\ref{sec: on Bloch Band Theory}, we formulate the non-Bloch band theory, analytically constructing the generalized Brillouin zone (GBZ) and defining a non-Bloch winding number that correctly captures the presence and absence of edge modes. Finally, we present our conclusions and outlook in Sec.~\ref{sec: Conclusions and Outlook}.

\begin{figure}
    \centering
    \includegraphics[width = \linewidth]{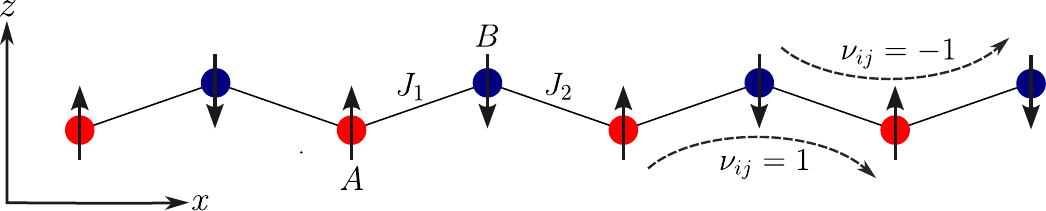}
    \caption{Schematic diagram of the antiferromagnetic chain. The spins interact with nearest neighbors by Heisenberg exchange interactions $J_1,\, J_2$. The spins interact to next nearest neighbour by DMI. The sign of DMI ($\nu_{ij} = \pm 1$) changes alternatively as shown. Each spin has an easy-axis anisotropy energy $\kappa$ (not shown) with the easy-axis directed along $z-$direction. The up and down spins are denoted by $A$ and $B$, respectively.}
    \label{fig:1}
\end{figure}

\section{Model Hamiltonian}
\label{sec: Model Hamiltonian}

We consider a spin chain with inversion symmetry about the bond centers, oriented along the $x$-axis. The minimal Hamiltonian includes Heisenberg exchange, symmetry-allowed Dzyaloshinskii--Moriya interactions (DMI), and easy-axis anisotropy,
\begin{equation}
    \mathcal{H} = \sum_{\langle i,j\rangle}J_{ij}\mathbf{S}_i\cdot \mathbf{S}_j 
    + D\sum_{\langle\langle i,j\rangle\rangle}\nu_{ij}\hat z\cdot \mathbf{S}_i\times \mathbf{S}_j 
    -\kappa\sum_i S_{iz}^2,
    \label{eq: full hamiltonian in 1d}
\end{equation}
where the first term describes the Heisenberg exchange interaction between nearest-neighbor spins. The exchange interaction is taken to be anisotropic, with \(J_{ij}>0\), corresponding to an antiferromagnetic system. To avoid double counting, we take all the summations over distinct pairs of sites. The second term is the Dzyaloshinskii--Moriya interaction (DMI) between next-nearest neighbors (NNN), where \(\nu_{ij}=\pm 1\) alternating between NNN bonds as shown in Fig.~\ref{fig:1}. This interaction induces canting of the spins. A nearest-neighbor DMI is forbidden due to inversion symmetry about the bond centers. Similarly, a uniform NNN DMI with $\nu_{ij} = 1$ is also forbidden by the same symmetry. The third term represents an easy-axis anisotropy along z-axis, which stabilizes the N\'eel state.

In the absence of the DMI $(D=0)$, the classical ground state is a collinear antiferromagnet with spins aligned antiparallel along the easy $z-$axis (N\'eel state). For small values of DMI \((0 < D \ll J_{ij})\), N\'eel state remains stable but for large values, the ground state is spin-spiral state. In the regime when the collinear antiferromagnet is the ground state, we bosonize the theory using the Holstein-Primakoff (HP) transformation~\cite{khomskii2010} to study the low energy excitation. Referring to Fig.~\ref{fig:1}, we take the HP transformations for $A$ sublattice as
\begin{subequations}
\begin{align}
    S_i^+=\sqrt{2S}a_i; \quad S_i^-=\sqrt{2S}a_i^\dagger;\quad S_{iz} = (S-a_i^\dagger a_i),
\end{align}
and for $B$ sublattice,
\begin{align}
    S_i^+=\sqrt{2S}b_i^\dagger;\quad S_i^-=\sqrt{2S}b_i;\quad S_{iz} = - (S-b_i^\dagger b_i).
\end{align}
\label{eq: HP Transformations}
\end{subequations}
Here, $S$ denotes the maximum value of $S_z$ at each site. The operators $a_i$ ($a_i^\dagger$) and $b_i$ ($b_i^\dagger$) are bosonic annihilation (creation) operators describing spin deviations at site $i$. Expressing the Hamiltonian Eq.~\eqref{eq: full hamiltonian in 1d} in terms of these bosonic operators, we obtain
{
\setlength{\jot}{2pt}
\begin{align}
    \mathcal{H}/S =&\,\, J_1\sum_{s=1}^N \left(a_sb_s + a_s^\dagger b_s^\dagger+ a_s^\dagger a_s +b_s^\dagger b_s \right) \notag\\
    +& J_2\sum_{s=1}^{N-1} \left(a_{s+1}b_{s} + a^\dagger_{s+1}b^\dagger_{s}+a^\dagger_{s+1}a_{s+1}+b^\dagger_{s}b_{s} \right) \notag \\
    +& (-iD) \sum_{s=1}^{N-1}\left(a^\dagger_s a_{s+1} + b^\dagger_s b_{s+1} - a^\dagger_{s+1} a_s - b^\dagger_{s+1} b_s \right) \notag\\
    +& 2\kappa \sum_{s=1}^N \left(a^\dagger_s a_s + b^\dagger_s b_s \right),
    \label{eq: finite Hamiltonian}
\end{align}
}
where, $N$ is the number of unit cells, each consisting of two sublattices $A$ and $B$. For a periodic chain, in the Nambu basis, $\psi_k = (a_k\,\,\, b_k\,\,\, a_{-k}^\dagger\,\,\, b_{-k}^\dagger)^T$, the Hamiltonian can be written as, $ \mathcal{H} = \frac{S}{2}\sum_k\psi^\dagger_k H(k) \psi_k $ with
\begin{equation}
    H(k) = \begin{pmatrix}
        A(k) & 0 & 0 & B(-k) \\
        0 & A(k) & B(k) & 0 \\
        0 & B(-k) & A(-k) & 0 \\
        B(k) & 0 & 0 & A(-k)
    \end{pmatrix}.
    \label{eq: first quantized Hamiltonian}
\end{equation}
Here, we defined $A(k) = J_1 + J_2 - 2D\sin(2kd) + 2\kappa$, $B(k) = J_1 + J_2 e^{-2ikd}$ and $d$ is nearest neighbor distance.

In general, this Hamiltonian cannot be diagonalized by a unitary transformation, since the upper and lower half components of the Nambu basis satisfy commutation relations with different signs~\cite{xiao2009}. To address this, one constructs a dynamic matrix associated with the Hamiltonian, whose eigenvalues determine the spectrum, while the corresponding eigenvectors define the transformation matrix~\cite{xiao2009, Colpa1978, Rohit2023} (for completeness, a derivation is provided in Appendix~\ref{App: Dynamic derivation}). The dynamic matrix is given by
\begin{equation}
    D_\text{BdG}(k) = I_-H(k),
    \label{eq: dynamicMatrixPeriodic}
\end{equation}
where, $I_- = \sigma_z \otimes I_2$ and $I_2$ is the $2 \times 2$ identity matrix. In general, $D_\text{BdG}(k)$ is a non-Hermitian matrix; more specifically, it is pseudo-Hermitian, satisfying $\eta D_\text{BdG}^\dagger(k) \eta^{-1} = D_\text{BdG}(k)$ with $\eta = \sigma_z \otimes I_2$. This property ensures that the spectrum is either entirely real or that complex eigenvalues appear in conjugate pairs.

For a periodic chain, the eigenvalues of $D_\text{BdG}$ are always real for any value of $D$. For sufficiently large $D$, the lower magnon branch develops a minimum that softens to zero energy at a specific momentum. The vanishing of the magnon gap at a finite momentum signals that the N\'eel state has ceased to be the correct ground state, the system instead favors a spin-spiral order. The Holstein--Primakoff description around the N\'eel state therefore breaks down beyond this point. The critical DMI strength at which this transition occurs can be obtained analytically by requiring the minimum of the magnon spectrum to touch zero (Appendix~\ref{App: Instability of Neel State}),
\begin{equation}
    D_c = \frac{\sqrt{\mathcal{A} + \sqrt{\mathcal{A}^2 - 4J_1^2 J_2^2}}}{2\sqrt{2}},
    \label{eq: Dc}
\end{equation}
where $\mathcal{A} = 4\kappa(J_1 + J_2 + \kappa) + 2J_1 J_2$. For the parameters used in Fig.~\ref{fig:3}(b) ($J_1 = J_2 = 1$, $\kappa = 0.05$), this yields $D_c \approx 0.68$.

\begin{figure}
    \centering
    \includegraphics[width = \linewidth]{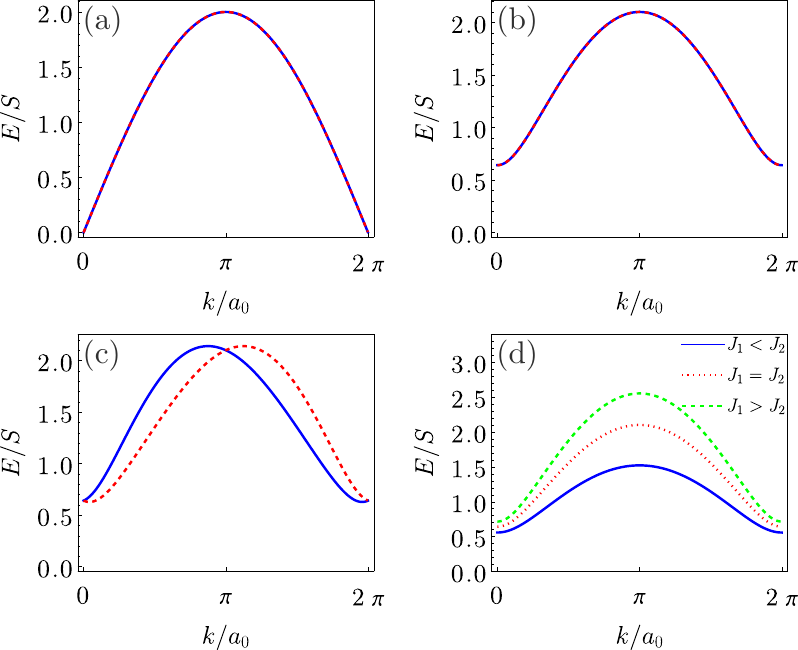}
    \caption{
    Magnon dispersion for different parameter values.
    (a) For $D = 0$, $\kappa = 0$, and $J_1 = J_2 = 1$, the system reduces to the Heisenberg model, exhibiting two degenerate gapless Goldstone modes. It is shown as overlapping blue solid and red dashed curves in the figure. Here, $a_0 = 2d$ is the lattice constant.
    (b) Introducing finite anisotropy ($\kappa = 0.05$) opens a gap at zero energy, with magnitude $2\sqrt{J_1 \kappa + J_2 \kappa + \kappa^2}$.
    (c) For finite Dzyaloshinskii--Moriya interaction ($D = 0.1$, $\kappa = 0.05$), the degeneracy between the two magnon branches is lifted.
    (d) The qualitative features remain unchanged for different exchange couplings; only the energy scales are modified. Here, we fix $D = 0$, $\kappa = 0.05$ and $J_2 = 1$, while vary $J_1 = 0.5, 1, 1.5$.
    }
    \label{fig:2}
\end{figure}

The dynamic matrix has particle hole symmetry \cite{Kawabata2019}, $T_- D^*_\text{BdG}(k) T_-^{-1} = -D_\text{BdG}(-k)$
with $T_- = \sigma_x \otimes I_2$. Consequently, for every eigenvalue $E(k)$, there exists a corresponding eigenvalue $-E(-k)$, when the spectrum is real. Since magnons are bosonic excitations, we restrict our attention to the positive-energy branches, which define the magnon dispersion. These are given by
\begin{equation}
    E_k^{\pm} = \sqrt{( J_1 + J_2 + 2 \kappa)^2 - \left| \left(J_1 + J_2 e^{-2ikd} \right)\right|^2 } \pm 2D \sin 2kd.
    \label{eqn: particle bands}
\end{equation}
The dispersion  for various parameter values is shown in Fig.~\ref{fig:2}. In absence of DMI and for $\kappa = 0$, the system reduces to the familiar Heisenberg model, which exhibits gapless Goldstone modes as shown in  Fig.~\ref{fig:2}(a). The other parameters are taken as $J_1 = J_2 = 1$. The system hosts two degenerate modes, represented by the overlapping blue solid and red dashed curves. For a finite value of $\kappa$, the energy gaps out from zero energy by an amount  $2\sqrt{J_1\kappa + J_2 \kappa + \kappa^2}$ as shown in Fig.~\ref{fig:2}(b). In the presence of DMI, this degeneracy is lifted, leading to a splitting of the two modes, as illustrated in Fig.~\ref{fig:2}(c). The qualitative behavior remains unchanged for different choices of $J_1$ and $J_2$ as shown in Fig.~\ref{fig:2}(d). Here, we present the dispersion for different $J_1$ values ($J_1 = 0.5, 1, 1.5$), while keeping $D = 0, \kappa = 0.05$ and $J_2 = 1$ fixed.

Throughout this work, we set $J_2 = 1$ unless otherwise stated.

\section{Finite System with Open Boundary}
\label{sec: Finite System with Open Boundary}

\begin{figure*}
    \centering
    \includegraphics[width=\linewidth]{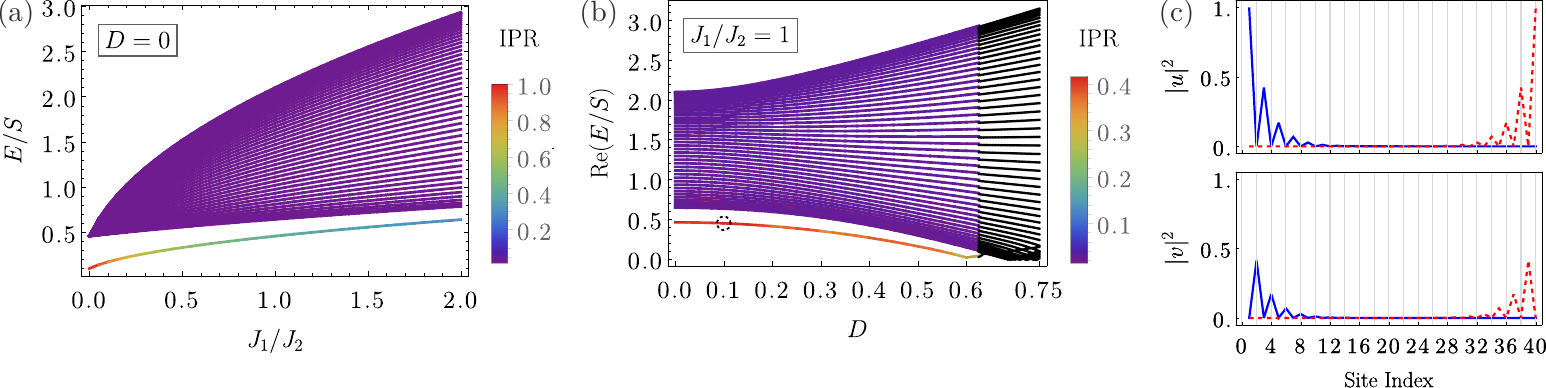}
    \caption{The energy spectrum and wavefunction profiles for a finite system under open boundary conditions.
    (a) Energy spectrum as a function of $J_1/J_2$, with $D = 0$ and $\kappa = 0.05$ fixed. The color scale represents the inverse participation ratio (IPR). Edge modes are present throughout the parameter range shown.
    (b) Energy spectrum as a function of $D$, with $J_1/J_2 = 1$ fixed. The edge states persist for finite $D$, with only a shift in the overall energy scale. For sufficiently large $D$, the edge states merge into the bulk continuum and the spectrum develops complex eigenvalues, signaling the onset of instability; this occurs around $D_c^f \approx 0.63$. Beyond this point, the spectrum is plotted in black.
    (c) Spatial profiles of the edge-mode wavefunctions for $D = 0.1$ (marked by the black dashed circle in panel (b)). Each eigenvalue in panel (b) is doubly degenerate; panel (c) shows the corresponding pair of states. The blue solid and red dashed lines represent the two degenerate edge states, where the particle~(hole) component of one state has support on sublattice $A$~($B$), while the particle~(hole) component of the other has support on sublattice $B$~($A$). The system size for panels (a) and (b) is 50 unit cells. For panel (c), a smaller system size of 20 unit cells is considered to emphasize localization on a specific sublattice at each edge.
    }
    \label{fig:3}
\end{figure*}

We now consider a finite system with open boundary conditions. The chain is terminated such that it contains an integer number of unit cells. The spectrum is obtained from the positive eigenvalues of the dynamic matrix. The spectrum and a representative wave function profile is shown in Fig.~\ref{fig:3}.  Each eigenstate of the dynamic matrix is written in the Nambu basis as $\psi = (u, v)$, where $u$ and $v$ denote the particle and hole amplitudes, respectively, with each being a vector of length $2N$ over the sublattice sites. The spectrum is color coded using the inverse participation ratio (IPR), computed using only the particle component~\cite{VermaPark2024},
\begin{equation}
\mathrm{IPR}
= \frac{\sum_{i=1}^{2N} \left| u_i \right|^{4}}{
    \left(\sum_{i=1}^{2N} \left| u_i \right|^{2} \right)^{2}},
\end{equation}
where $u_i$ is the $i$th element of the particle amplitude $u$, and the sum runs over all $2N$ sublattice sites. A high IPR value indicates a localized state.

In Fig.~\ref{fig:3}(a), we plot the energy as a function of $J_1/J_2$. The other parameters are chosen as $D = 0$ and $\kappa = 0.05$. In addition to the bulk states, we observe edge localized states that are well separated from the bulk spectrum. These edge states persist throughout the entire parameter range, with no special feature occurring at $J_1/J_2 = 1$. This behavior contrasts with that of a conventional SSH chain, where a topological phase transition occurs near equal values of the hopping amplitudes~\cite{asboth2016short}. In the SSH chain, zero-energy edge modes are protected by chiral symmetry of the bulk Hamiltonian. Here, however, the edge modes occur at finite energy, and their wave functions involve mixed particle-hole amplitudes with sublattice selectivity. Their origin lies in the bosonic BdG boundary problem. A finite value of $\kappa$ is essential, since for $\kappa = 0$ the bulk gap closes at zero energy. The spectrum is also doubly degenerate. In Fig.~\ref{fig:3}(b), we show the effect of DMI by plotting the energy as a function of $D$ for fixed $J_1/J_2 = 1$. The edge states persist for small values of $D$ ($D < 0.6$). Around $D_c^f \approx 0.63$, the spectrum begins to develop complex eigenvalues. Beyond this transition point, the real part of the spectrum is shown in black. We note that the onset of instability in the finite chain, $D_c^f \approx 0.63$, is lower than the critical value $D_c \approx 0.68$ obtained for the periodic chain from the bulk gap closure condition $E^-(k^*) = 0$. The two thresholds correspond to distinct mechanisms: in the periodic chain, the instability is driven by the softening of a bulk magnon mode to zero energy, whereas in the finite chain it is triggered by the coalescence of edge and bulk states at an exceptional point~\cite{Ohashi2020} at finite energy. For a finite system, the DMI does not lift the degeneracy of the spectrum. In Fig.~\ref{fig:3}(c), we plot the wave function profiles of two degenerate edge states for a representative value of parameters (indicated by the black dashed circle in panel (b) at $D = 0.1$). The blue solid and red dashed lines represent the two degenerate states. For each edge state, the particle component is localized on one sublattice while the hole component on the other. This sublattice polarization is a direct consequence of the block-diagonal structure of the dynamical matrix, where each block does not mix the sublattices.

\section{Bulk Topological Invariant}
\label{sec: Bulk Winding Number}
\subsection{Winding Number}
A bulk winding number can be defined as a topological invariant for the periodic system. Following Ref.~\cite{Ohashi2020}, we introduce a Hermitian winding number. In the absence of DMI, the dynamic matrix possesses a sublattice symmetry $S\, D_\text{BdG}(k)\, S^{-1} = -D_\text{BdG}(k)$ with $S = \sigma_x \otimes I_2$. Together with pseudo-Hermiticity, this allows one to construct a Hermitian matrix
\begin{equation}
    D_0(k) = [D_\text{BdG}(k) + D_\text{BdG}^\dagger(k)]/2
\end{equation}
which inherits a sublattice symmetry $\tilde{S}\, D_0(k)\, \tilde{S}^{-1} = -D_0(k)$ with $\tilde{S} = \sigma_y \otimes I_2$.
We can define a winding number as we do for Hermitian systems
\begin{equation}
    w = \frac{1}{2\pi}\text{Im}\Big[\int_{-\pi}^\pi  dk \, \partial_k \text{ln}\big[\text{det } \tilde q(k) \big]\Big]
\end{equation}
where, $\tilde q(k)$ is defined by
\begin{equation*}
    U_{\tilde S}^\dagger D_0(k) U_{\tilde S}
    =
    \begin{pmatrix}
        0 & \tilde q(k) \\
        \tilde q^\dagger (k) & 0
    \end{pmatrix}
\end{equation*}
with $U_{\tilde S}$ being the unitary matrix that diagonalizes $\tilde S$. For the present model $\tilde q (k) = -2 (J_1 + J_2 + 2 \kappa)\, I$, which is real and independent of $k$. The winding number therefore vanishes, $w = 0$, for all parameter values.

\subsection{\texorpdfstring{$\mathbb{Z}_2$}{Z2} Topological Invariant}

In absence of DMI, the system is inversion symmetric, satisfying $PD_\text{BdG}(k)P^{-1} = D_\text{BdG}(-k),$ with $P = I_2 \otimes \sigma_x$. In this case, one can define a $\mathbb{Z}_2$ topological invariant~\cite{Ohashi2020}. Let $|u_n^R(k)\rangle$ denote the right eigenstate of $D_\text{BdG}(k)$ corresponding to the eigenvalue $E_n(k)$. For the $n$th particle band, we define the biorthogonal Berry connection as
\begin{equation}
    A_{nn}(k) = i\langle u^R_n(k)|I_-|\partial_k u^R_n(k)\rangle.
\end{equation}
We define the $\mathbb{Z}_2$ topological invariant as
\begin{equation}
    (-1)^\nu = \exp\left\{i\int_{-\pi}^\pi dk\,\sum_{n = 1}^2 A_{nn}(k)\right\}.
\end{equation}
This invariant is also found to be trivial, i.e., $\nu = 0$ (Appendix~\ref{App: topInvariant}).

A more complete understanding can be obtained from the 38 fold symmetry classification introduced in Ref.~\cite{Kawabata2019, Gong2018} which serves as a non-Hermitian generalization of the celebrated Altland-Zirnbauer symmetry~\cite{Altland1997,ryu2010topological}. In the absence of DMI ($D=0$), the system has particle-hole symmetry, $C_- D^T_\text{BdG}(k) C_-^{-1} = -D_\text{BdG}(-k)$ with $C_- = \sigma_y \otimes I_2$ and time-reversal symmetry, $D^*_\text{BdG}(k) = D_\text{BdG}(-k)$. It thus belongs to class CI$(\eta_{+-})$ in the 38-fold classification, where $\eta_{+-}$ means that the pseudo-Hermiticity operator commutes with time-reversal but anti-commutes with particle-hole symmetry. In presence of DMI, the time reversal symmetry is broken. The system falls into class C$(\eta_-)$ where, $\eta_-$ means the pseudo-Hermiticity operator anti-commutes with particle-hole symmetry. Our system exhibits a real line gap ($L_r$). According to the non-Hermitian topological classification, one-dimensional systems in class CI$(\eta_{+-})$ or C$(\eta_-)$ with a real line gap do not support a nontrivial topological invariant~\cite{Kawabata2019}. This explains the absence of a bulk topological characterization. Nevertheless, the finite chain under OBC supports edge states as shown in Fig.~\ref{fig:3}. This mismatch between the trivial Bloch invariant and the nontrivial boundary spectrum motivates the non-Bloch analysis and is discussed later.

\section{Boundary Perturbation and Tunable Edge Modes}
\label{sec: Boundary Perturbation and Tunable Edge Modes}

For a finite system under open boundary conditions, the absence of exchange bonds at the ends of the chain naturally modifies the local environment of the boundary sites. This leads to a shift in onsite energies at boundary sites. Motivated by this, we introduce a tunable boundary perturbation that allows manipulation of the onsite energies at the edges. Such boundary terms can also be engineered externally, as discussed later. We consider an onsite energy term added to the Hamiltonian in Eq.~\eqref{eq: finite Hamiltonian}:
\begin{equation}
    \mathcal{H}_\text{pert} = \epsilon(a_1^\dagger a_1 + b_N^\dagger b_N).
    \label{eq: generalisedBC}
\end{equation}
The resulting dispersion as a function of the boundary energy $\epsilon$ is shown in Fig.~\ref{fig:4}(a)--(c). In Fig.~\ref{fig:4}(a), the parameters are fixed to $J_1 = 0.5$, $D = 0$, and $\kappa = 0.05$, and the color scale represents the IPR. We observe that the energies of the edge states increase approximately linearly with $\epsilon$. Beyond a critical value of $\epsilon$ (indicated by vertical dashed blue lines), the edge states merge into the bulk bands and subsequently reemerge from the opposite side of the spectrum. For finite $D$, the energies of the states are slightly modified; however, the overall behavior remains unchanged. The disappearance and reappearance of the edge states are not restricted to the regime $J_1 < J_2$. In Fig.~\ref{fig:4}(b), we present the same analysis for $J_1 = 1$, where the points at which the edge states merge with and separate from the bulk shift, while the qualitative behavior remains the same. In Fig.~\ref{fig:4}(c), we consider $J_1 = 1.5$.

The boundary potential introduced in Eq.~\eqref{eq: generalisedBC} can be realized experimentally by locally modifying the environment at the edges of the spin chain. Two possible ways to realize such a boundary term are, applying a local Zeeman field at the boundary sites or modifying the magnetic anisotropy at the edges. For magnetic fields applied only at the two ends, the Zeeman term reads
\begin{equation}
    \mathcal{H}_z = -\gamma \mathcal{B} (S_{1z} - S_{Nz}),
\end{equation}
where $\gamma$ is the gyromagnetic ratio and $\mathcal{B}$ is the magnitude of the applied magnetic field. Using the Holstein--Primakoff transformations Eq.~\ref{eq: HP Transformations}, the Zeeman Hamiltonian becomes
\begin{equation}
\mathcal{H}_z = \gamma \mathcal{B} (a_1^\dagger a_1 + b_N^\dagger b_N) + \text{const}.
\end{equation}
Alternatively, a boundary contribution can arise from a local modification of the anisotropy,
\begin{equation}
\mathcal{H}_K = -K_{1N} \left(S_{1z}^2 + S_{Nz}^2\right),
\end{equation}
where $K_{1N}$ denotes the anisotropy strength at the edge sites. Substituting the Holstein--Primakoff expressions and keeping only the quadratic terms in the bosonic operators within the linear spin-wave approximation, we obtain
\begin{equation}
\mathcal{H}_K = 2K_{1N}S (a_1^\dagger a_1 + b_N^\dagger b_N) + \text{const}.
\end{equation}

\section{Non-Bloch Band Theory}
\label{sec: on Bloch Band Theory}

The conventional winding number defined for a periodic chain fails to account for the edge states. The sensitivity of the edge modes to boundary perturbations indicates the need for a boundary-sensitive invariant. To address this, we formulate a winding number for the finite system. In this section, we use the methodology of the Non-Bloch theory developed in Ref.~\cite{Yokomizo2021, Guo2021}. The idea is to solve the finite system analytically. Although translational invariance is preserved in the bulk, it is broken at the boundaries. We solve the resulting finite-difference equations in the bulk and impose appropriate boundary conditions. Since the system is finite, momentum is no longer a good quantum number. However, one can introduce a generalized (complex) momentum, which allows us to establish an energy--generalized-momentum relation. The detailed calculations are presented in the Appendix~\ref{App: nonBloch}.

The bulk matrix can be obtained from the dynamic matrix Eq.~\eqref{eq: dynamicMatrixPeriodic} by making the substitution $e^{2ikd} \to z^{-1}$ and $e^{-2ikd} \to z$. For the analytic construction we set $D = 0$. The resulting bulk matrix $M_B$ can be block-diagonalized by an appropriate basis transformation. In the basis $\{u_A,\, v_B,\, v_A,\, u_B\}$, it takes the form
\begin{equation}
M_B =
    \begin{pmatrix}
    M_{B1} & 0 \\
    0 & -M_{B1}
    \end{pmatrix}, \,\,
    M_{B1} =
    \begin{pmatrix}
    \mathcal{F} & J_1+\frac{J_2}{z} \\
    -(J_1+zJ_2) & -\mathcal{F}
    \end{pmatrix}.
    \label{eq: bulkMatrixnonBlochmainText}
\end{equation}
We defined $\mathcal{F} = J_1 + J_2 + 2\kappa$ for convenience. Both blocks yield identical characteristic equation. Considering the first block, the characteristic equation is
\begin{align}
    \text{Det}[M_{B1} - E] &= 0  \notag\\
    J_1 J_2 z^2 +(E^2 - 2 J_1 J_2 -4 J_1 \kappa - & 4 J_2 \kappa -4 \kappa^2)z + J_1 J_2 = 0,
    \label{eq: genEngymainText}
\end{align}
which is a quadratic equation in $z$. The two solutions $z_1$ and $z_2$ satisfy $z_1 z_2 = 1$. The corresponding bulk eigenvectors can be obtained from each block. For first block, we obtain
\begin{equation}
    v_B = \frac{E - \mathcal{F}}{J_1 + \frac{J_2}{z}}u_A,
\end{equation}
while for the second
\begin{equation}
    v_A = \frac{E - \mathcal{F}}{J_1+zJ_2}u_B.
\end{equation}
The general solution is then given by
\begin{equation}
    \psi_s = c_1 z_1^s \psi_1(z_1) + c_2 z_2^s \psi_1(z_2) + c_3 z_1^s \psi_2(z_1) + c_4 z_2^s \psi_2(z_2).
    \label{eq: genSolnmainText}
\end{equation}
where $c_i$ are constants determined by the boundary conditions. In writing this form, we have excluded the degenerate case $z_1 = z_2$.
\begin{figure*}
    \centering
    \includegraphics[width=\linewidth]{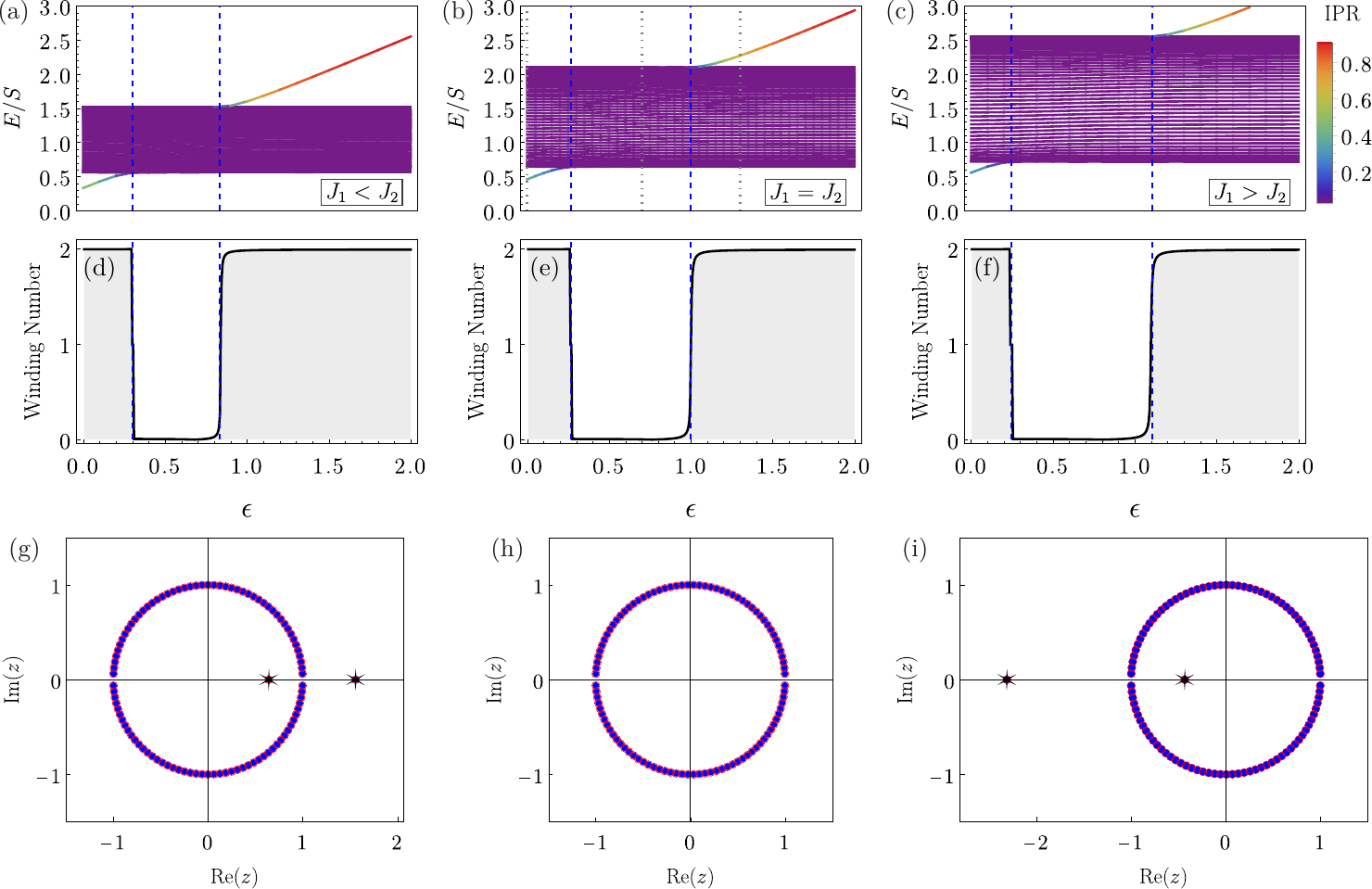}
    
    \caption{
    Energy spectrum, winding number, and generalized Brillouin zone (GBZ).
    (a)--(c) Energy as a function of $\epsilon$ for $J_1 = 0.5, 1, 1.5$, respectively, with $J_2 = 1, D = 0, \kappa = 0.05$ fixed. The color scale represents the inverse participation ratio (IPR). In all cases, edge states appear under open boundary conditions (OBC) \textit{i.e.} at $\epsilon  = 0$. As $\epsilon$ is increased, these states merge into the bulk spectrum and subsequently re-emerge from the other side of the bands.
    (d)--(f) Shows the corresponding non-Bloch winding number, capturing the appearance and disappearance of edge modes (also indicated by vertical blue dashed lines).
    In (g)--(i), we display GBZ for three representative values of $\epsilon$ ($\epsilon = 0, 0.7, 1.3$, respectively), as indicated by vertical dotted grey lines in panel (b). Blue dots denote numerical results, while red empty circles represent analytical solutions, showing excellent agreement. For $\epsilon = 0$, the bulk states lie on the unit circle, while isolated points correspond to edge states (indicated by `$*$'). As $\epsilon$ increases ($\epsilon = 0.7$), these states merge into the unit circle, and for larger $\epsilon$ ($\epsilon = 1.3$), they re-emerge on the opposite side.}
    \label{fig:4}
\end{figure*}
We solve the system with perturbed boundary conditions Eq.~\eqref{eq: generalisedBC}. Substituting the general solution Eq.~\eqref{eq: genSolnmainText} into the boundary conditions, we obtain
\begin{equation}
    \begin{pmatrix}
        P_1(z_1) & P_1(z_2) & 0 & 0 \\
        Q_1(z_1) & Q_1(z_2) & 0 & 0 \\
        0 & 0 & P_2(z_1) & P_2(z_2) \\
        0 & 0 & Q_2(z_1) & Q_2(z_2)
    \end{pmatrix}
    \begin{pmatrix}
        c_1 \\ c_2 \\ c_3 \\ c_4
    \end{pmatrix} = 0
    \label{eq: boundaryMmainText}
\end{equation}
where,
\begin{subequations}
\begin{align}
    P_1(z) &= z (J_2 - \epsilon) + \frac{J_2(E - \mathcal{F})}{J_1 + \frac{J_2}{z}} \\
    Q_1(z) &= - J_2 z^{N+1} + \frac{z^N(\epsilon - J_2)(E - \mathcal{F})}{J_1 + \frac{J_2}{z}} \\
    P_2(z) &= z(\epsilon - J_2) - \frac{J_2(E - \mathcal{F})}{J_1 + z J_2} \\
    Q_2(z) &= J_2 z^{N + 1} + \frac{z^N(J_2 - \epsilon)(E - \mathcal{F})}{J_1 + z J_2}.
\end{align}
\end{subequations}
We denote the coefficient matrix in Eq.~\eqref{eq: boundaryMmainText} as $M_b$ and denoting each block as $M_{b1},M_{b2}$. A nontrivial solution requires $\det[M_b] = 0$.
Considering the first block $ \text{det}[M_{b1}] = 0 $, after some algebraic manipulations, one get
\begin{align}
    J_1 z ^{2N} (\epsilon + J_2z -J_2)^2 - J_1(J_2 + \epsilon z &- J_2 z)^2 \notag \\
     - J_2 z (z^{2N} - 1)(4J_2\kappa &- 4\epsilon\kappa - \epsilon^2) = 0
     \label{eq: solnAnalyticmainText}
\end{align}
which determines the allowed values of the generalized momentum $z$. The corresponding energies are obtained from \eqref{eq: genEngymainText},
\begin{equation}
    E^2 = 2 J_1 J_2 + 4 (J_1 + J_2) \kappa + 4 \kappa^2 -J_1 J_2 \left(z + \frac{1}{z}\right).
    \label{eq: energyAnalyticMainText}
\end{equation}
In Fig.~\ref{fig:4}(g)--(i), we plotted the GBZ for three different values of $\epsilon$ (indicated by grey dotted vertical lines in panel (b)). The blue dots denote numerically obtained results i.e.~energies obtained by exact diagonalization of the real space dynamic matrix, is put in Eq.~\eqref{eq: genEngymainText} to obtain the generalized momenta, $z$. Red empty circles denote the results obtained analytically i.e.~ solving Eq.~\eqref{eq: solnAnalyticmainText}. The two disjoint generalized momenta away from GBZ shown in these figures specify the edge states (indicated by `$*$'). In Fig.~\ref{fig:4}(g), for $\epsilon = 0$, we see the bulk states lie on the unit circle, while isolated points correspond to edge states. Their localization length can be obtained by using $\xi = - 1 / \log |z^*|$. In Fig.~\ref{fig:4}(h), As $\epsilon$ increases ($\epsilon = 0.7$), these states merge into the unit circle, and for larger $\epsilon$ ($\epsilon = 1.3$) as in Fig.~\ref{fig:4}(i), they re-emerge on the opposite side.

\subsection{Non-Bloch Topological invariant}
We define a winding number on generalized Brillouin zone (GBZ). Following Ref.~\cite{Verma2024GBZ, VermaPark2024}, we introduce the function $h_+(z_1, z_2) = P_1(z_2)/P_1(z_1)$
and define the corresponding winding number as
\begin{equation}
    W_+ = \frac{1}{2\pi i}\oint
    \frac{1}{h_+(z)}\frac{d h_+(z)}{dz}\,dz
\end{equation}
where we have used the relation $z_1 z_2 = 1$. The integration goes over the contour formed by the generalized momenta, $z$ characterizing the bulk states, which, in this case, corresponds to the unit circle. Similarly, we define $h_-(z_1, z_2) = Q_1(z_1)/Q_1(z_2)$
with the associated winding number
\begin{equation}
    W_- = \frac{1}{2\pi i}\oint
    \frac{1}{h_-(z)}\frac{d h_-(z)}{dz}\,dz.
\end{equation}
Finally, the non-Bloch winding number is defined as $W_{\text{nB}} = (W_+ - W_-)/2$. The quantization of $W_\pm$ follows from the structure of the boundary matrix. One can construct an equivalent transformed boundary matrix, $\widetilde{M}_{b1}$ that shares the same eigenvectors as $M_{b1}$ but exhibits an emergent chiral symmetry:
\begin{equation}
    \widetilde{M}_{b1} = \begin{pmatrix}
        0 & h_+(z_1, z_2) \\
        h_-(z_1, z_2) & 0
    \end{pmatrix}
\end{equation}
which satisfies $\{\widetilde{M}_{b1}, \sigma_z \} = 0$. It can also be shown that $W_+ = -W_-$, by the following way. From the condition $\det[M_{b1}] = 0$, we obtain
\begin{equation}
    P_1(z_1) Q_1(z_2)\left[1 - h_+(z_1, z_2) h_-(z_1, z_2)\right] = 0.
\end{equation}
Assuming $P_1(z_1) Q_1(z_2) \neq 0$ on the contour, this implies $h_+(z_1, z_2) = 1/h_-(z_1, z_2)$.
This implies $W_+ = -W_-$, so that the non-Bloch winding number reduces to $W_{\text{nB}} = W_+$. In Fig.~\ref{fig:4}(d)--(f), we plot the winding number as a function of $\epsilon$. In Fig.~\ref{fig:4}(a)--(c), we presented the energy spectrum as a function of $\epsilon$, illustrating the appearance and disappearance of edge states. The corresponding winding numbers are shown in panels (d)--(f), respectively, directly below each spectrum. The threshold values is indicated by the vertical dashed blue lines. The invariant correctly characterizes the appearance and disappearance of edge states in all cases.

The above analysis has been carried out in the absence of DMI ($D = 0$). For finite DMI, the bulk matrix $M_B$ remains block-diagonal in the chosen basis. The effect of DMI is to shift both diagonal elements of each block by the same term, $iD(z^{-1} - z)$. However, the characteristic equation for each block becomes quartic in $z$, yielding four generalized momenta. Following the formalism of Ref.~\cite{Yokomizo2019}, the GBZ can be obtained numerically by ordering the solutions with increasing magnitude, $|z_1| \leq |z_2| \leq |z_3| \leq |z_4|$, and identifying the contour via the condition $|z_2| = |z_3|$. The quartic order of $z$, however, complicates an analytic determination of the GBZ, and accordingly the evaluation of the non-Bloch winding number in this regime becomes more involved.

\section {Conclusion and Outlook}
\label{sec: Conclusions and Outlook}

We studied a one-dimensional antiferromagnetic spin chain incorporating Heisenberg exchange interaction, Dzyaloshinskii-Moriya interaction (DMI), and easy-axis anisotropy. Such a model can be realized in layered van der Waals materials. Within the linear spin-wave approximation, the system belongs to the CI($\eta_{\pm}$) symmetry class in the absence of DMI, and to the C($\eta_{-}$) class when DMI is present, according to the 38-fold classification of non-Hermitian systems \cite{Kawabata2019}. In one dimension, these symmetry classes do not support any nontrivial bulk topological invariant. Nevertheless, the model exhibits edge states in a finite system under open boundary conditions. These states originate from the inequivalence between bulk and boundary sites induced by lattice termination. This termination effectively modifies the local onsite energies at the edges, giving rise to a generalized boundary condition. We find that the energies of these edge states can be tuned by introducing an on-site energy perturbation at the boundary sites. By varying this perturbation, the edge states initially present under open boundary conditions merge into the bulk spectrum and subsequently re-emerge as the perturbation strength is increased. To understand this, we employ the Non-Bloch band theory framework and define a winding number for the finite system. For analytical tractability, we focus on the case of vanishing DMI and show that the winding number is quantized to 2 when edge states are present. It drops to zero when the edge states merge with the bulk and returns to 2 when they reappear. A complete analytical treatment for the case of finite DMI is left for future work.

We now comment on the broader applicability of our framework. While our analysis has focused on antiferromagnetic chains, the emergence of a bosonic Bogoliubov-de Gennes (BdG) structure is not restricted to this setting. Similar quadratic forms also arise in ferromagnetic systems with biaxial/triaxial anisotropies or pseudo-dipolar interactions~\cite{Wei_2022}, where anomalous terms appear in the spin-wave Hamiltonian. As a result, the corresponding magnon dynamics in such systems are likewise governed by non-Hermitian dynamic matrices. This suggests that the boundary-induced phenomena and non-Bloch characterization developed here can, in principle, be extended to a broader class of magnetic systems. A detailed exploration of these extensions is left for future work.

Beyond this, it is also natural to consider extensions of the minimal model studied here. Incorporating additional interactions would allow for a more realistic description of van der Waals systems and may alter the symmetry class. In situations where the symmetry class does not support a bulk topological invariant yet edge states exist in the finite system, it remains an open question whether a non-Bloch winding number can still characterize the emergence of edge states.

Finally, the edge modes identified here are sublattice-polarized and remain well separated from the bulk spectrum over a broad parameter range, requiring a finite easy-axis anisotropy and a stable N\'eel ground state. Their sublattice polarization and spatial localization make them promising candidates for experimental detection. In particular, spatially resolved probes such as scanning nitrogen-vacancy center magnetometry~\cite{Thiel2019}, which offers nanoscale resolution, and nonlocal magnon transport measurements~\cite{Xing2019}, could be used to distinguish boundary and bulk contributions. Related magnon transport and nonlinear magnon-response phenomena in van der Waals antiferromagnets provide additional routes for probing magnetic excitations and interaction effects~\cite{PhysRevB.107.245426,PhysRevB.108.165412}. Brillouin light scattering, which is well suited for probing magnon spectra in thicker or layered samples, may also provide a complementary route to detect these modes. An interesting direction for future work is to extend the present analysis to higher-dimensional systems, where the boundary modes identified here may evolve into corner~\cite{PhysRevB.97.241405} or higher-order modes accessible to such experimental techniques.

\section{ACKNOWLEDGMENTS}
R.M. is grateful for the useful communication with Ran Cheng (University of California, Riverside) and Flore Kunst (Max Plank). S.V. acknowledges support from the Brain Pool Program funded by the Ministry of Science and ICT through the National Research Foundation of Korea (RS-2025-25446099). R.M.\ acknowledges support from a postdoctoral fellowship from the Alexander von Humboldt Foundation. S.D. acknowledges the institute fellowship from IIT Kanpur. A.K. acknowledges funding from ANRF (SERB) through projects ANRF/ARG/2025/002460/PS
and ANRF/ARGM/2025/002682/TS.

\appendix
\section{Derivation of Dynamic Matrix}
\label{App: Dynamic derivation}
To diagonalize the Hamiltonian Eq.~\ref{eq: first quantized Hamiltonian}, one can do a basis transformation $\psi_k = T_k \phi_k$.
Demanding the new creation/annihilation operators satisfy bosonic commutation rules, $T_k$ matrix needs to satisfy paraunitarity condition
\begin{equation}
    T_kI_-T^\dagger_k = I_-,
    \label{eqAppend: paraunitarityk}
\end{equation}
where, $I_- = \sigma_z \otimes I_2$ with $I_2$ is the $2 \times 2$ identity matrix. The matrix $T_k$ diagonalizes the Hamiltonian
\begin{equation}
    T^\dagger_k H(k) T_k = \begin{pmatrix}
        E_k & \\
        & E_{-k}
    \end{pmatrix},
    \label{eqAppend: first Hamiltoniank}
\end{equation}
where, $E_k$ is a $2 \times 2$ diagonal matrix of eigenvalues. From \eqref{eqAppend: paraunitarityk} and \eqref{eqAppend: first Hamiltoniank}, one can show
\begin{align}
    I_- H(k) T_k &= T_k I_- \begin{pmatrix}
        E_k & \\
        & E_{-k}
    \end{pmatrix} \\
    D_\text{BdG}(k) T_k &= T_k \begin{pmatrix}
        E_k & \\
        & -E_{-k}
    \end{pmatrix}.
    \label{app: eigenvalue_Matrix}
\end{align}
Thus, the eigenvalues of $D_\text{BdG}(k) \equiv I_-H(k)$ determine the eigenvalues of the Hamiltonian and the right eigenvectors determine the paraunitary matrix $T_k$. The right hand eigenvalue matrix in Eq.~\eqref{app: eigenvalue_Matrix} takes the specific form due to the particle-hole symmetry of $D_\text{BdG}(k)$.  The eigenvalues of the dynamic matrix are given by
\begin{equation}
    E_k^{\pm} = - 2D \sin(2kd) \pm \sqrt{\mathcal{F}^2 - \left|(J_1 + J_2 e^{-2ikd}) \right|^2 },
\end{equation}
where $\mathcal{F} = J_1 + J_2 + 2\kappa$. The spectrum is doubly degenerate. Reflecting negative branch about $E=0$ yields the particle bands Eq.~\eqref{eqn: particle bands}.

\section{Instability of the N\'eel State}
\label{App: Instability of Neel State}
The instability of the N\'eel state is determined by the DMI value at which the lower magnon branch touches zero energy. We consider the lower branch obtained from Eq.~\eqref{eqn: particle bands},
\begin{align}
E_{-}(\phi)
= \sqrt{\mathcal{A} - 2J_1J_2\cos\phi} - 2D\sin\phi .
\end{align}
For convenience, we defined $\phi = 2kd$ and introduced $\mathcal{A} = 4\kappa(J_1 + J_2 + \kappa) + 2J_1J_2$.
The extrema of the dispersion satisfy
$\frac{\partial E_-}{\partial \phi}=0,$
which gives
\begin{align}
\frac{J_1J_2\sin\phi}{\sqrt{\mathcal{A} - 2J_1J_2\cos\phi}}
-2D\cos\phi =0 .
\label{eqAppend: interim1}
\end{align}
The N\'eel state becomes unstable when the minimum of the spectrum reaches zero energy, i.e.,
$E_- = 0 .$
This condition implies
\begin{equation}
\sqrt{\mathcal{A} - 2J_1J_2\cos\phi} = 2D\sin\phi .
\label{eqAppend: interim2}
\end{equation}
Using relations Eq.~\eqref{eqAppend: interim1} and Eq.~\eqref{eqAppend: interim2}, one obtains a quadratic equation in $4D^2$
\begin{align}
(4D^2)^2 - \mathcal{A}(4D^2) + J_1^2J_2^2 = 0 .
\end{align}
Solving for $D$ yields
\begin{equation}
D = \frac{\sqrt{\mathcal{A} + \sqrt{A^2 - 4J_1^2J_2^2}}}{2\sqrt{2}}.
\end{equation}
At this point the magnon gap closes and the N\'eel state becomes unstable.

\section{\texorpdfstring{$\mathbb{Z}_2$}{Z2} Topological Invariant}
\label{App: topInvariant}
The idea is to continuously deform the dynamic matrix while keeping the spectral gap open. In this way, if the eigenvectors become momentum independent, the Berry connection vanishes.

Dynamic matrix for a periodic chain Eq.~\eqref{eq: dynamicMatrixPeriodic}, in absence of DMI takes the form
\begin{equation}
    D_\text{BdG}(k) = \begin{pmatrix}
        \mathcal{F}I_2 & Q(k) \\
        -Q(k) & -\mathcal{F}I_2
    \end{pmatrix}
\end{equation}
where \(\mathcal{F} = J_1 + J_2 + 2\kappa\), \(I_2\) is the \(2 \times 2\) identity matrix, and
\[
Q(k) =
\begin{pmatrix}
0 & B(-k) \\
B(k) & 0
\end{pmatrix}, \quad
B(k) = J_1 + J_2 e^{-2ikd}.
\]
We introduce a continuous deformation:
\begin{equation}
D^\lambda_\text{BdG}(k) =
\begin{pmatrix}
\mathcal{F}I_2 & \lambda Q(k) \\
- \lambda Q(k) & - \mathcal{F}I_2
\end{pmatrix}, \qquad 0 \le \lambda \le 1.
\end{equation}
The corresponding eigenvalues are
\begin{equation}
E = \pm \sqrt{\mathcal{F}^2 - \lambda^2 |B(k)|^2}.
\end{equation}
Since $\mathcal{F} > |B(k)|$ for all $k$ when $\kappa > 0$, the gap remains open for all $\lambda \in [0,1]$. Thus, \(D^\lambda_\text{BdG}(k)\) is adiabatically connected to
\begin{equation}
D^0_\text{BdG}(k) =
\begin{pmatrix}
\mathcal{F}I_2 & 0 \\
0 & -\mathcal{F}I_2
\end{pmatrix},
\end{equation}
without closing the gap. In this limit, the eigenvectors are momentum independent, and therefore the Berry connection
\begin{equation}
A_{nn}(k) = i \langle u_n^R(k) | I_- | \partial_k u_n^R(k) \rangle
\end{equation}
vanishes. Consequently, the $\mathbb{Z}_2$ Topological Invariant, $\nu$:
\begin{equation}
(-1)^\nu = \exp\!\left(i \int_{-\pi}^{\pi} dk \sum_{n=1}^2 A_{nn}(k)\right)
\end{equation}
is zero.

\section{Non Bloch Band Theory}
\label{App: nonBloch}
In this section, we present the non Bloch theory in detail following Ref.~\cite{Yokomizo2021}. For the analytic construction we set $D = 0$. The Hamiltonian in real space Eq.~\eqref{eq: finite Hamiltonian}, can be written as
\begin{equation}
    \mathcal{H} = \frac{S}{2} \psi^\dagger H \psi,
\end{equation}
where, $\psi = \begin{pmatrix}
    a_1 & b_1 & \dots & a_N & b_N & a^\dagger_1 & b^\dagger_1 & \dots & a^\dagger_N & b^\dagger_N
\end{pmatrix}^T$ is in Numbu basis. To diagonalize this Hamiltonian one can do a basis transformation, $\psi = T \phi$. Demanding the new creation/annihilation operators satisfy bosonic commutation rules, $T$ matrix needs to satisfy paraunitarity condition~\cite{xiao2009}
\begin{equation}
    TI_-T^\dagger = I_-,
    \label{eqAppend: paraunitarity}
\end{equation}
where, $I_- = \text{diag} (I, -I)$ with $I$ is identity matrix of dimension $2N$. The matrix $T$ diagonalizes the Hamiltonian
\begin{equation}
    T^\dagger H T = M,
    \label{eqAppend: first Hamiltonian}
\end{equation}
where $M$ is diagonal. From \eqref{eqAppend: paraunitarity} and \eqref{eqAppend: first Hamiltonian}, one can show
\begin{equation}
    I_- H T = T I_-M
    \label{eqAppend: dynamicMatrix}
\end{equation}
which defines the dynamic matrix $D_{\text{BdG}} \equiv I_- H$. The eigenvalues of $D_{\text{BdG}}$ determine the eigenvalues of the Hamiltonian and the right eigenvectors determine the paraunitary matrix $T$. Owing to the paraunitary constraint, the components of $T$ are not all independent and can be parametrized in the form
\begin{equation}
    T = \begin{pmatrix}
        A & B \\
        B^* & A^*
    \end{pmatrix}.
\end{equation}
The matrices $A$ and $B$ are constructed from the particle and hole components of the eigenvectors, respectively. Explicitly, they take the form
\begin{equation}
    A =
    \begin{pmatrix}
        u^{(1)}_{A1} & u^{(1)}_{B1} & \cdots & u^{(1)}_{AN} & u^{(1)}_{BN} \\
        u^{(2)}_{A1} & u^{(2)}_{B1} & \cdots & u^{(2)}_{AN} & u^{(2)}_{BN} \\
        \vdots & \vdots & \ddots & \vdots & \vdots \\
        u^{(2N)}_{A1} & u^{(2N)}_{B1} & \cdots & u^{(2N)}_{AN} & u^{(2N)}_{BN}
    \end{pmatrix},
\end{equation}
and
\begin{equation}
    B =
    \begin{pmatrix}
        v^{(1)}_{A1} & v^{(1)}_{B1} & \cdots & v^{(1)}_{AN} & v^{(1)}_{BN} \\
        v^{(2)}_{A1} & v^{(2)}_{B1} & \cdots & v^{(2)}_{AN} & v^{(2)}_{BN} \\
        \vdots & \vdots & \ddots & \vdots & \vdots \\
        v^{(2N)}_{A1} & v^{(2N)}_{B1} & \cdots & v^{(2N)}_{AN} & v^{(2N)}_{BN}
    \end{pmatrix},
\end{equation}
where $(k)$ labels the $k$th eigenvector. Owing to the particle-hole symmetry of $D_{\text{BdG}}$, the matrix $I_- M$ takes the form $\mathrm{diag}(\mathbf{E}, -\mathbf{E})$, where $\mathbf{E}$ is a diagonal matrix of dimension $2N$ containing the positive eigenvalues. Evaluating Eq.~\eqref{eqAppend: dynamicMatrix} for the $s$th unit cell, we obtain the bulk equations
\begin{subequations}
\begin{align}
    \mathcal{F}\, u_{A,s} + J_2 v_{B,s-1} + J_1 v_{B,s} &= E u_{A,s}\\
    \mathcal{F}\, u_{B,s} + J_1 v_{A,s} + J_2 v_{A,s+1} &= E u_{B,s} \\
    -J_2 u_{B,s-1} - J_1 u_{B,s} - \mathcal{F}\, v_{A,s} &= E v_{A,s} \\
    -J_1 u_{A,s} - J_2 u_{A,s+1} - \mathcal{F}\, v_{B,s} &= E v_{B,s}
\end{align}
\end{subequations}
for $s = 2, 3, \dots, N-1$. We defined $\mathcal{F} = J_1+J_2+2\kappa$ for convenience. To solve these equations, we take the ansatz, $(u_{\alpha,s}, v_{\alpha,s}) \propto z^s (u_{\alpha}, v_{\alpha})$, $\alpha \in \{A,B\}$. The bulk equations become
\begin{equation}
    M_B
    \begin{pmatrix}
        u_A \\ u_B \\ v_A \\ v_B 
    \end{pmatrix} = 
    E \begin{pmatrix}
        u_A \\ u_B \\ v_A \\ v_B 
    \end{pmatrix}
\end{equation}
with
\begin{equation}
    M_B = \begin{pmatrix}
        \mathcal{F} & 0 & 0 & J_1 + \frac{J_2}{z} \\
        0 & \mathcal{F} & J_1 + z J_2 & 0 \\
        0 & - J_1 -\frac{J_2}{z} & - \mathcal{F} & 0 \\
        -J_1 - z J_2 & 0 & 0 & - \mathcal{F}
    \end{pmatrix}.
\end{equation}
This is the matrix, one would get by replacing $e^{2ikd} \to z^{-1}$ and $e^{-2ikd} \to z$ in Eq.~\eqref{eq: dynamicMatrixPeriodic}, and is given by Eq.~\eqref{eq: bulkMatrixnonBlochmainText} in the main text. The bulk matrix $M_B$ can be made block diagonalised by a basis change. In $\{u_A,\, v_B,\, v_A,\, u_B\}$, $M_B$ becomes
\begin{equation}
M_B =
    \begin{pmatrix}
    M_{B1} & 0 \\
    0 & -M_{B1}
    \end{pmatrix}, \,
    M_{B1} =
    \begin{pmatrix}
    \mathcal{F} & J_1+\frac{J_2}{z} \\
    -(J_1+zJ_2) & -\mathcal{F}
    \end{pmatrix}.
\end{equation}
Both blocks yield identical characteristic equation. Considering the first block, the characteristic equation is
\begin{align}
    \text{Det}[M_{B1} - E] &= 0  \notag\\
    J_1 J_2 z^2 +(E^2 - 2 J_1 J_2 -4 J_1 \kappa - & 4 J_2 \kappa -4 \kappa^2)z + J_1 J_2 = 0.
    \label{eqAppend: characteristic}
\end{align}
This is a quadratic equation in $z$ and the roots satisfy $z_1 z_2 = 1$. For first block $M_{B1}$, the bulk eigenvectors obey
\begin{equation}
    v_B = \frac{E - \mathcal{F}}{J_1 + \frac{J_2}{z}}u_A.
\end{equation}
The eigenvector is (fixing $u_A = 1$)
\begin{equation}
    \psi_1(z) = \begin{pmatrix}
        1 \\ 0 \\ 0 \\ \frac{E - \mathcal{F}}{J_1 + \frac{J_2}{z}}
    \end{pmatrix}.
\end{equation}
Similarly for the other block, we get
\begin{equation}
    v_A = \frac{E - \mathcal{F}}{J_1+zJ_2}u_B.
\end{equation}
The eigenvector is (fixing $u_B = 1$)
\begin{equation}
    \psi_2(z) = \begin{pmatrix}
        0 \\ 1 \\ \frac{E - \mathcal{F}}{J_1+zJ_2} \\ 0
    \end{pmatrix}.
\end{equation}
The general solution will be
\begin{equation}
    \psi_s = c_1 z_1^s \psi_1(z_1) + c_2 z_2^s \psi_1(z_2) + c_3 z_1^s \psi_2(z_1) + c_4 z_2^s \psi_2(z_2),
    \label{eqAppend: genSoln1}
\end{equation}
where $c_i$ are constants that need to be determined from the boundary conditions. Explicitly, the above form gives
\begin{subequations}
\begin{align}
    u_{A,s} &= c_1 z_1^s + c_2 z_2^s \\
    v_{A, s} &= c_3 z_1^s + c_4 z_2^s \\
    v_{B, s} = c_1 z_1^s \frac{E - \mathcal{F}}{J_1 + \frac{J_2}{z_1}} &+ c_2 z_2^s \frac{E - \mathcal{F}}{J_1 + \frac{J_2}{z_2}} \\
    u_{B, s} = c_3 z_1^s \frac{E - \mathcal{F}}{J_1 + z_1 J_2} &+ c_4 z_2^s \frac{E - \mathcal{F}}{J_1 + z_2 J_2}.
\end{align}
\label{eqAppend: genSol}
\end{subequations}
In writing \eqref{eqAppend: genSoln1}, we have excluded the case, $z_1 = z_2$.

We solve the equations for the perturbed system. In presence of the boundary perturbation \eqref{eq: generalisedBC}, the boundary conditions are given by
\begin{subequations}
\begin{align}
    (J_2-\epsilon) u_{A,1} + J_2 v_{B,0} &= 0 \\
    -J_2 u_{B,0} + (-J_2 + \epsilon) v_{A,1} &= 0 \\
    (J_2 - \epsilon) u_{B,N} + J_2 v_{A,N+1} &= 0 \\
    -J_2 u_{A,N+1} + (-J_2+\epsilon) v_{B,N} &= 0.
\end{align}
\end{subequations}
We can put the general solutions Eq.~\eqref{eqAppend: genSol} above to obtain
\begin{equation}
    \begin{pmatrix}
        P_1(z_1) & P_1(z_2) & 0 & 0 \\
        Q_1(z_1) & Q_1(z_2) & 0 & 0 \\
        0 & 0 & P_2(z_1) & P_2(z_2) \\
        0 & 0 & Q_2(z_1) & Q_2(z_2)
    \end{pmatrix}
    \begin{pmatrix}
        c_1 \\ c_2 \\ c_3 \\ c_4
    \end{pmatrix} = 0
    \label{eq: boundaryM}
\end{equation}
where,
\begin{subequations}
\begin{align}
    P_1(z) &= z (J_2 - \epsilon) + \frac{J_2(E - \mathcal{F})}{J_1 + \frac{J_2}{z}} \\
    Q_1(z) &= - J_2 z^{N+1} + \frac{z^N(\epsilon - J_2)(E - \mathcal{F})}{J_1 + \frac{J_2}{z}} \\
    P_2(z) &= z(\epsilon - J_2) - \frac{J_2(E - \mathcal{F})}{J_1 + z J_2} \\
    Q_2(z) &= J_2 z^{N + 1} + \frac{z^N(J_2 - \epsilon)(E - \mathcal{F})}{J_1 + z J_2}.
\end{align}
\end{subequations}
We define the coefficient matrix in Eq.~\eqref{eq: boundaryM} as $M_b$ and denoting each block as $M_{b1},M_{b2}$. For non-trivial solutions, one needs $\text{det}[M_b] = 0$. Considering the first block $ \text{det}[M_{b1}] = 0 $, after some algebraic manipulations, one get
\begin{align}
    J_1 z ^{2N} (\epsilon + J_2z -J_2)^2 - J_1(J_2 + \epsilon z &- J_2 z)^2 \notag \\
     - J_2 z (z^{2N} - 1)(4J_2\kappa &- 4\epsilon\kappa - \epsilon^2) = 0.
\end{align}
Solutions of this equation gives the generalized momenta. The energies are obtained from \eqref{eqAppend: characteristic},
\begin{equation}
    E^2 = 2 J_1 J_2 + 4 (J_1 + J_2) \kappa + 4 \kappa^2 -J_1 J_2 \left(z + \frac{1}{z}\right).
\end{equation}
This energy equation is presented as Eq.~\eqref{eq: energyAnalyticMainText} in the main text.

\bibliographystyle{apsrev4-2}
\bibliography{references}

\end{document}